\documentclass[titlepage,12pt,twoside]{article}
\usepackage{amssymb,epsfig,pslatex}
\usepackage{cite}  
\usepackage{float}     
\usepackage{wrapfig}  
\usepackage{hhline} 
\usepackage{dcolumn}
\restylefloat{figure}

\makeatletter
\def\@sect#1#2#3#4#5#6[#7]#8{\ifnum #2>\c@secnumdepth
  \def\@svsec{}\else
  \refstepcounter{#1}\edef\@svsec{\csname the#1\endcsname.\hskip0.5em}\fi
  \@tempskipa #5\relax
  \ifdim \@tempskipa>\z@
    \begingroup
      #6\relax 
      \@hangfrom{\hskip #3\relax\@svsec}{\interlinepenalty \@M #8\par}%
    \endgroup 
    \csname #1mark\endcsname{#7}\addcontentsline 
      {toc}{#1}{\ifnum #2>\c@secnumdepth \else 
        \protect\numberline{\csname the#1\endcsname}\fi #7}%
  \else 
    \def\@svsechd{#6\hskip #3\@svsec #8\csname #1mark\endcsname 
      {#7}\addcontentsline{toc}{#1}{\ifnum #2>\c@secnumdepth \else 
        \protect\numberline{\csname the#1\endcsname}\fi #7}}%
  \fi \@xsect{#5}} 
\@addtoreset{equation}{section} 
\makeatother 
\renewcommand\thesection{\arabic{section}} 
\renewcommand\theequation{\ifnum \value{section}>0 
 \thesection.\arabic{equation}%
\else 
\arabic{equation}%
\fi} 

\renewcommand{\thefootnote}{\small\fnsymbol{footnote}} 
 
\begin{document} 
\begin{titlepage} 
  \begin{flushright} 
    PITHA 08/01 
  \end{flushright}         
\vspace{0.01cm} 
\begin{center} 
{\LARGE {\bf  Determining the CP parity of Higgs bosons\\[4pt]
    at the LHC\\[14pt] 
 in their $\tau$ decay channels}}  \\ 
\vspace{2cm} 
{\large{ 
\bf Stefan Berge\footnote{\tt berge@physik.rwth-aachen.de}, \, Werner Bernreuther\footnote{\tt breuther@physik.rwth-aachen.de}, \, J\"org Ziethe 
}} 
\par\vspace{1cm} 
Institut f\"ur Theoretische Physik, RWTH Aachen, 52056 Aachen, Germany\\ 
\par\vspace{1cm} 
{\bf Abstract}\\ 
\parbox[t]{\textwidth} 
{If neutral Higgs bosons will be discovered at the 
CERN Large Hadron Collider (LHC) then an important subsequent issue will 
be the investigation of their $CP$ nature. Higgs boson decays into 
$\tau$ lepton pairs are particularly suited in this 
respect. Analyzing the three charged pion 
decay modes of the $\tau$ leptons and taking 
expected measurement uncertainties at the LHC into account, 
we show that the $CP$ properties of 
a Higgs boson can be pinned down with appropriately chosen 
observables, 
provided that sufficiently large event numbers will eventually be available. 
} 
\end{center} 
\vspace*{2cm} 
 
PACS number(s): 11.30.Er, 12.60.Fr, 14.80.Bn, 14.80.Cp  
\end{titlepage} 
%
%
\oddsidemargin=-2mm
\setcounter{footnote}{0} 
\renewcommand{\thefootnote}{\arabic{footnote}} 
\setcounter{page}{1} 
 
{\it \bf Introduction:} \quad  
The major physics goal of the upcoming experiments 
 at the CERN Large Hadron Collider (LHC) is the exploration of 
the hitherto unknown mechanism of electroweak gauge symmetry breaking 
which, in the context of the standard model of particle physics (SM) 
and many of its extensions, is tantamount to searching for Higgs bosons, 
spin-zero and electrically neutral resonances with masses of (a few) hundred 
GeV (see, e.g., \cite{Djouadi:2005gi,Accomando:2006ga} for reviews). 
If (one or several types of) Higgs bosons are found then the next issue 
will be the determination of their properties -- in particular their 
parity $(P)$ and   charge conjugation times 
parity $(CP)$ quantum numbers, respectively,  which  
yield important information about the dynamics of these particles. 
While the 
SM  Higgs boson is parity-even, 
SM extensions also predict parity-odd state(s) or, if the  
Higgs boson dynamics violates $CP$, states of undefined $CP$ parity with Yukawa 
couplings both to scalar and pseudoscalar quark and lepton currents. 
Higgs sector $CP$ violation is a fascinating possibility, especially 
in view of its potentially enormous impact 
on an important issue of the  physics of the early universe,  
namely baryogenesis 
\cite{Cohen:1993nk}. 
These new interactions can 
be searched for at the upcoming generation of colliders  in 
several ways (see \cite{Accomando:2006ga} for a collection of recent results).  
The decays of Higgs bosons to  $\tau^- \tau^+$ leptons and/or -- if the 
Higgs bosons are heavy enough -- to  
$t \bar t$ quarks are particularly suited in this respect 
 \cite{Bernreuther:1993df,Bernreuther:1997af,Bernreuther:1998qv,Dell'Aquila:1988fe,Chang:1993jy,Kramer:1993jn,Grzadkowski:1995rx}. 
In this letter we show that, at the LHC, the $CP$ nature of a neutral Higgs boson  
can be pinned down with appropriately  
chosen observables in its $\tau^- \tau^+$ decay channel where the 
$\tau$ decay into three charged pions, provided  that sufficiently  
large event numbers will eventually be  available. 
 
The analysis of this letter applies to any neutral Higgs boson $h_j$ with  
flavor-diagonal couplings to quarks and leptons $f$ (with mass 
$m_f$) 
\begin{equation} 
{\cal L}_{Y}=-(\sqrt{2}G_F)^{1/2} \sum_{j,f} m_f(a_{jf} \bar{f}f + 
b_{jf} \bar{f}i\gamma_5 f)\,h_j \,\, ,  
\label{bbzyukl} 
\end{equation} 
where $G_F$ is the Fermi constant and $a_{jf}$ and $b_{jf}$ are the reduced  
scalar and pseudoscalar Yukawa couplings respectively, which depend on the  
parameters of the (effective) Higgs potential of the respective  model.  
In the SM, $a_{f}=1$ and $b_{f}=0$. As far as SM extensions are concerned we 
consider here, for definiteness, models with two Higgs doublets, such as 
the non-supersymmetric type II models and the minimal supersymmetric SM extension 
(MSSM) (see, e.g., \cite{Djouadi:2005gi,Accomando:2006ga}). 
These models contain  three physical neutral 
Higgs fields $h_j$ in the mass basis. 
If Higgs sector $CP$ violation (CPV) is negligibly small then the fields 
$h_j$  describe two scalar states $h, H$ ($b_{f}=0)$ 
and a pseudoscalar $A$ $(a_{f}=0)$.  
In the case of Higgs sector CPV, the $h_j$ 
have non-zero couplings\footnote{They can be parameterized 
in terms of the ratio of the Higgs field 
vacuum expectation values $\tan\beta = {\rm v}_2/{\rm v}_1$ 
and a  $3\times 3$ orthogonal  
matrix that describes the mixing of the neutral  
spin-zero CP eigenstates \cite{Bernreuther:1992dz}.} 
$a_{jf}$ and $b_{jf}$ to quarks and leptons which lead to $CP$-violating 
effects in $h_j\to f{\bar f}$ already at the Born level \cite{Bernreuther:1993df}. 
This is in contrast to the couplings to $W^+W^-$ and to $ZZ$ boson pairs of such 
a state of undefined $CP$ parity. 
At  the Born level, only the $\rm{CP}=+1$ component of $h_j$ couples to 
$W^+W^-$ and to $ZZ$. The coupling of the pseudoscalar component of 
$h_j$ -- if there is any -- to $W^+W^-$ and to $ZZ$ is likely to be very small as it 
must be induced by quantum fluctuations. Thus, the observation of Higgs boson production in weak 
vector boson fusion $W^+W^-, ZZ \to h_j$ or of the decay channels  
$ h_j \to W^+W^-, ZZ$ would tell us that  $h_j$ has a significant  
scalar component. However, the question would remain  
whether or not $h_j$ is a pure $J^{PC}=0^{++}$ state 
or if it has a significant pseudoscalar component, too\footnote{ 
Nevertheless, it can be checked with appropriate correlations whether 
or not there is a sizeable effective pseudoscalar coupling to weak 
vector bosons \cite{Dell'Aquila:1985ve}.} 
. This can be answered by investigating 
\nopagebreak{the $\tau$ decay channel of this particle}. \\ \\ 
{\it \bf $\tau$ spin observables:} \quad  
In the following, we choose the generic notation $\phi$ for 
any of the neutral Higgs bosons $h_j$ discussed above. 
The observables discussed here 
 for determining   
the $CP$ quantum number of $\phi$ in its  decay channel $\phi \to \tau^- \tau^+$  
may be applied to any Higgs production process. 
At the LHC, this includes the gluon and gauge boson 
fusion processes $gg \to \phi$ and $q_i q_j \to \phi  q'_i q'_j$,  
respectively, and the associated production  
$t {\bar t} \phi$ or $b {\bar b} \phi$ of a light Higgs boson. 
We consider the following semi-inclusive 
reactions  
\begin{equation} 
i \, \to \, \phi \, + \, X \, \to \,  \tau^-({\bf k}_\tau, \alpha) \, + \,  
\tau^+({\bf k}_{\bar \tau}, \beta) \, + \, X \,, 
\label{bbzreac1} 
\end{equation} 
where $i$ is some partonic initial state, 
${\bf k}_\tau$ and 
${\bf k}_{\bar \tau} = -{\bf k}_\tau$ are the 3-momenta of $\tau^-$ and $\tau^+$ 
in the $\tau \bar \tau$ zero-momentum frame (ZMF), and $\alpha,\beta$ are spin  
labels. Here we make use of the fact that at colliders  
polarization and spin correlation effects are both measurable 
and reliably predictable for tau leptons. 
 
Let us assume that experiments at the LHC will discover 
a neutral boson  
resonance in  a reaction of the type (\ref{bbzreac1})  and a sufficiently 
large sample will eventually be accumulated. 
The spin of $\phi$ may be inferred from the polar angle distribution 
of the tau leptons. Suppose the outcome of this analysis is that $\phi$ is a 
spin-zero (Higgs) particle. One would next like to determine 
its Yukawa coupling(s), and specifically like to know whether  
$\phi$ is a scalar, a pseudoscalar, or 
a state with undefined $CP$ parity. 
This can be investigated by using the following $CP$-even and -odd  
observables involving the spins of $\tau, {\bar \tau}$, 
 and we emphasize that all of them should be used. 
The correlation 
resulting from projecting the spin of  $\tau$ onto the spin of $\bar \tau$, 
\begin{equation} 
{S} \, = \, {{\mathbf s}_\tau} \cdot {{\mathbf s_{\bar \tau}}} \, , 
\label{bbzo1} 
\end{equation} 
is the best choice for discriminating  between a 
$CP= \pm 1$ state  \cite{Bernreuther:1997af}. Here $ {{\mathbf s}_\tau}, {{\mathbf s_{\bar \tau}}}$ denote the spin operators of 
$\tau^-$ and $\tau^+$, respectively. This can be understood as follows. 
If $\phi$ is a scalar ($J^{PC}=0^{++}$) then 
${\tau} {\bar{\tau}}$ is in a $^3P_0$ state, and  
$\langle{\bf s}_{\tau} \cdot {\bf s}_{\bar{\tau}}\rangle = 1/4$.  
If $\phi$ is a pseudoscalar ($J^{PC}=0^{-+}$) 
then  ${\tau} {\bar {\tau}}$ is in a $^1S_0$ state  
and $\langle{\bf s}_{\tau} \cdot {\bf s}_{\bar{\tau}}\rangle = -3/4$, which is  
strikingly different from the scalar case.  
For general couplings (\ref{bbzyukl}) one gets $\langle{S}\rangle 
=(a_\tau^2-3b_\tau^2)/(4a_\tau^2+4b_\tau^2)$, 
 using that $m_\phi\gg m_\tau$ \cite{Bernreuther:1997af}. 
 
If $\gamma_{CP}^\tau \equiv - a_\tau b_\tau \neq 0$, 
the Yukawa interactions of $\phi$ to $\tau$ leptons are not $CP$-invariant.  
This leads to 
$CP$-violating effects  in the  reactions (\ref{bbzreac1}). 
For an  unpolarized initial state $i$,  
 a  general kinematic analysis of (\ref{bbzreac1}) yields   
the following result 
\cite{Bernreuther:1993df,Bernreuther:1998qv}. 
If $C$-violating interactions do not matter in 
(\ref{bbzreac1}) then  
${\cal L}_{Y}$ (which is  $C$-invariant, but $P$- and $CP$-violating)   
induces  two types of CPV effects: 
a  $CP$-odd $\tau^- \tau^+$ 
spin-spin correlation   and a $CP$-odd $\tau$ polarization asymmetry which  
correspond to the  observables 
\begin{equation} 
{S}_{CP} \, = \,{\bf \hat k}_\tau \cdot( {{\mathbf s}_\tau}  
\times {{\mathbf s_{\bar \tau}}} ) \, , 
\quad        {S}'_{CP} \, = \,{\bf \hat k}_\tau\cdot({{\mathbf s}_\tau} - {{\mathbf s_{\bar \tau}}}) \,. 
\label{bbzo23} 
\end{equation} 
Here,  ${\bf \hat k}_\tau  ={\bf k}_\tau/|{\bf k}_\tau|$  in the 
$\tau\bar \tau$ ZMF\footnote{Two more terms can appear in the 
squared matrix element of (\ref{bbzreac1}). They are 
obtained by replacing  ${\bf\hat k}_\tau  \to {\bf \hat p}$  
 in (\ref{bbzo23}),  where ${\bf \hat  p}$  is the direction of one of 
the colliding beams in $i$. However, 
for resonant $\phi$ production only the observables  
(\ref{bbzo23}) are of interest.}. 
The $CP$-odd and $T$-odd\footnote{Here $T$-even/odd refers to 
a naive $T$ transformation, i.e., reversal of momenta and spins only.} 
variable ${S}_{CP}$ measures a correlation of the spins  
of the $\tau^-$ and $\tau^+$  
transverse to their  directions of flight. A non-zero expectation value 
is generated already at tree level,  
$\langle{S}_{CP}\rangle =  - a_\tau b_\tau/(a^2_\tau + b_\tau^2)$ 
\cite{Bernreuther:1993df}, which can be as large as 0.5 in magnitude.  
The variable ${S}'_{CP}$ 
measures an asymmetry in the longitudinal polarization of the  
$\tau^-$ and $\tau^+$. 
As it is $CP$-odd but $T$-even, a non-zero  $\langle {S}'_{CP}\rangle$  
requires both 
 $ \gamma_{CP}^\tau \neq 0$ and a non-zero absorptive part of 
the respective scattering amplitude. 
 
{\it \bf (Multi) pion final states:} \quad The polarization and spin-correlation effects   
induced in the $\tau^- \tau^+$ sample from Higgs boson decay 
lead in turn, through the parity-violating weak decays of 
the $\tau$ leptons, to specific angular distributions  
and correlations in the respective final state which can be measured with 
appropriately constructed observables (see below).  
In order to obtain sufficient 
sensitivity to the $CP$ properties of the Higgs boson resonance, 
one should consider $\tau$ 
decay channels that have both  good to maximal $\tau$-spin 
 analyzing power and  allow  for the reconstruction of the $\tau$ 
decay vertex, i.e., the $\tau$ rest frame, which is essential for an efficient 
 helicity analysis. 
 
We recall  the $\tau$-spin analyzing power of 
the final state $a$ in the decay $\tau^-\to a + \nu_\tau$, that is, 
the coefficient $c_a$ in the distribution $\Gamma_a^{-1}d \Gamma_a/ d\cos \theta_a = 
(1+ c_a \,\cos\theta_a)/2$, where $\theta_a$ is the 
angle between the $\tau^-$ spin vector and the direction of $a$ in the 
$\tau^-$ rest frame (c.f., e.g., \cite{Stahl:2000aq}). $CP$ invariance   
which is, as known from experiments, 
a good symmetry in $\tau$ decays at the level of precision required here 
implies that  the $\tau$-spin analyzing power of $\bar a$ in  
$\tau^+\to {\bar a} + {\bar \nu}_\tau$ is  $c_{\bar a}= -c_a$. 
 
The $\pi^-$ channel is known 
to have maximal spin-analyzing power, $c_{\pi^-} =1$. In the case of, e.g.,  multi-pion decays of 
the $\tau$, this optimum analyzing power can also be  achieved if all the pions are 
observed and the dependence of the hadronic current on the pion momenta is known 
\cite{Rouge:1990kv,Davier:1992nw,Kuhn:1995nn}. The latter is obtained using empirically 
tested matrix elements and fits to measured distributions.  
 
In the following section 
we consider the decay $\tau^- \to \pi^- \pi^- \pi^+ \nu_\tau$ and 
the corresponding decay of $\tau^+$. As this decay proceeds to a large extent via the 
$a_1$ resonance, we use  $\tau^\mp \to a_1^\mp$ 
for the description of the three-pion final state.  The measured pion 
momenta in the laboratory frame allow, using known kinematic distributions  
\cite{Rouge:1990kv,Davier:1992nw,Stahl:2000aq}, the separation of the longitudinal 
$(a_{1L})$ and transverse $(a_{1T})$ helicity states of the $a_1$. This leads to  
an optimal spin analyzing power, $c=\pm 1$ for $a_{1L}$ and $a_{1T}$, respectively. 
Moreover, the  measured pion 
momenta yield the $\tau$ decay vertex and, in turn, the $\tau$ rest frame (see below). 
 
We thus investigate reactions of the following type:   
\begin{equation}  
p \, p  \, \to \, \phi \, + \, X \, \to \, 
\tau^-({\bf k}_\tau, \alpha) \, + \,  
\tau^+({\bf k}_{\bar \tau}, \beta) \, + \, X \, 
\to \, a({\bf q}_1) \,+ \, 
 {\bar b}({\bf q}_2) \,+ \, X \,, 
\label{bbzreac2} 
\end{equation} 
where  $a$ and  $\bar b$ denote here the $a_1^-$ and $a_1^+$ 
resonances, respectively. Without loss of generality, we consider  
the $a_{1L}^-$ and $a_{1L}^+$ states and comment on  
how the resulting distributions change when one or two $a_{1T}$ are involved. 
 The momenta of $\tau^-$ and $\tau^+$ 
in (\ref{bbzreac2}) are defined as above in the $\tau \bar\tau$ ZMF,  
while the momenta ${\bf q_1}$ 
and ${\bf q_2}$ refer to the $\tau^-$ and $\tau^+$ rest frames, respectively.  
 
Let's now come to the equivalents of the ${S}_i$ at the level 
of the final states $a, \bar b$. The spin correlation  
$\langle{S}\rangle$ leads to a non-isotropic distribution 
in $\cos\varphi$, where $\varphi =\angle({\bf q_1}, {\bf 
  q_2})$. If no phase space cuts are applied -- modulo cuts on the  
invariant mass $M_{\tau\bar \tau}$ of the $\tau$ pair -- this 
opening angle distribution  is  of the form  
\cite{Bernreuther:1997af}: 
\begin{equation} 
\frac{1}{\sigma_{a{\bar b}}} \frac{d \sigma_{a{\bar b}}}{d\cos \varphi} =   
\frac{1}{2} \left( 1 - D_{a{\bar b}} \, \cos\varphi \right) \, , 
\quad 
D_{a{\bar b}}  =  \frac{4}{3} \, c_a c_b \, \langle{\bf s}_{\tau}  
\cdot {\bf s}_{\bar{\tau}}\rangle \, . 
\label{bbzoadi} 
\end{equation} 
The  coefficients 
$D_{a_{1L}^- a_{1L}^+}=D_{a_{1T}^- a_{1T}^+}$ are $0.33$ and 
$-1$ for the channels  
$\phi \, (0^{++}) \to \tau \tau \to a_{1i} a_{1j}$ 
and  $\phi \, (0^{-+}) \to \tau \tau \to a_{1i} a_{1j}$, respectively, 
 if $ij=LL \,, TT$, while they change sign if  $ij=LT \,, TL$. 
Thus, for $ij=LL \,, TT$ 
the $a_1$  momenta ${\bf q}_1$, ${\bf q}_2$ are predominantly parallel 
in the case of a pseudoscalar $\phi$, 
while for a scalar $\phi$ they tend to be antiparallel.  
 
The equivalents of the $CP$-odd spin observables ${S}_{CP}$ and 
  ${S}'_{CP}$ at the level of the final states (\ref{bbzreac2}) 
are \cite{Bernreuther:1997af,Bernreuther:1998qv}: 
\begin{equation} 
{\cal O}_{CP} \, = \, ({{\hat{\mathbf k}}_\tau} -  
{\mathbf{\hat{k}}_{\bar \tau}}) 
\cdot({\mathbf{\hat{q}}_2} \times {\mathbf{\hat{q}}_1} )/2 \,\, ,  
\quad {\cal O}'_{CP} \, = \,  {{\hat{\mathbf k}}_\tau}\cdot  
{\mathbf{\hat{q}}_1} - 
{\mathbf{\hat{k}}_{\bar \tau}}  \cdot {\mathbf{\hat{q}}_2} \,\, .  
\label{bbzqcp} 
\end{equation} 
 
In general, if $b \neq a$ 
the averages of (\ref{bbzqcp})  should be taken for events
(\ref{bbzreac2}) plus 
the charge conjugated events ${\bar a} b$.
Asymmetries corresponding to (\ref{bbzqcp}) are: 
\begin{equation} 
A({\cal O}) \, = \, \frac{N_{a{\bar b}}({\cal O}>0) - N_{a{\bar b}}({\cal O}<0)} 
                    {N_{a{\bar b}}} \,\,,  
\label{bbzasq} 
\end{equation} 
where $N_{a{\bar b}}$ is the number of events in the reaction 
(\ref{bbzreac2}).   
If no phase-space cuts, besides cuts on   
$M_{\tau\bar\tau}$, are imposed then \cite{Bernreuther:1998qv} 
\begin{eqnarray} 
\langle {\cal O}_{CP} \rangle_{a{\bar b}} \, = \, -\frac{4}{9} c_a
c_{\bar b} \, \langle {S}_{CP} 
\rangle \, \, , \quad  
\langle {\cal O}'_{CP} \rangle_{a{\bar a}} \, = \, \frac{2}{3} c_a \, \langle {S}'_{CP} 
\rangle \, \, , \nonumber \\ 
A({\cal O}_{CP}) \, = \, \frac{9\pi}{16} \langle {\cal O}_{CP} \rangle_{a{\bar b}}  \, , \quad 
A({\cal O}'_{CP}) \, = \, \,\langle {\cal O}'_{CP} \rangle_{a{\bar a}}        \,\, . 
\label{bbzresa} 
\end{eqnarray} 
Here the relations involving ${\cal O}'_{CP}$ 
are given for simplicity for the diagonal channels $a \bar a$ only.  

The observable ${\cal O}_{CP}$ measures 
the distribution of the signed normal vector of  the plane spanned by  
${\bf q_1}, {\bf  q_2}$ with respect to the $\tau^-$ direction of 
flight. If $\gamma_{CP}^\tau \neq 0$ then this distribution is 
asymmetric.  If $\phi$ were an ideal mixture of a $CP$-even and -odd state, 
$|a_\tau|=|b_\tau|$, the asymmetry corresponding to ${\cal O}_{CP}$ would take 
the value $|A({\cal O}_{CP})| =0.4$ in the $a_{1i}a_{1j}$ channels $(i,j=L,T)$. 
As already mentioned, $\langle {\cal O}'_{CP} \rangle \neq 0$ also requires, besides 
$\gamma_{CP}^\tau \neq 0$, an absorptive part in the scattering 
amplitude $i \to \phi \to \tau^- \tau^+$, which is small in our 
analysis below. Therefore, 
we do not consider this observable any further. \\ \\ 
{\it \bf Results:} \quad  
For non-standard Higgs bosons $\phi$ and large  $\tan\beta$, the 
associated production with bottom 
quarks, $gg\to b\bar{b}\phi$, is considered to be the most promising 
mode in the $\phi \to \tau\bar{\tau}$ decay channel at the 
LHC \cite{:1999fr,Ball:2007zza}.  
The  results shown below are not only  
applicable to this production process, but also to 
gluon-gluon fusion, $gg\to\phi$, and vector boson fusion. The reason 
is that our  normalized distributions do not depend on the $\phi$ 
momentum  if no detector cuts are applied. Furthermore we show   
for $\phi\to\tau\bar{\tau}\to a_{1}\bar{a}_{1}$ 
that detector cuts have only a very small effect on these distributions for  
Higgs masses larger than $200$~GeV. Thus,  our results will not 
change significantly for  the different Higgs production modes  
or if initial-state higher-order QCD corrections are taken into account.  
We have, therefore,  computed in this analysis all distributions for a generic  
$2\to1$ Higgs boson production process at leading order.  
 
As emphasized above, the determination of  the distributions of 
$\cos\varphi$ and of the observable ${\cal O}_{CP}$ requires the reconstruction of 
the $\tau^\mp$ rest frames. For 
the decay channels (\ref{bbzreac2}), 
the $a_1^\mp$ momenta in the laboratory frame and  
the $\tau^\mp$ decay vertices can be obtained from the visible tracks  
of the three charged pions.  
The $\tau^\mp$ production, i.e., the Higgs production vertex,  can be 
reconstructed  
from the visible tracks of the charged particles/jets produced in association with 
the $\phi$ \cite{genn}. 
Using that, for each $\tau$,  
${\tilde k}_{\tau}^{\mu}={\tilde q}_{a_{1}}^{\mu}+{\tilde q}_{\nu}^{\mu}$, 
$m_{\tau}^{2}={\tilde E}_{\tau}^{2}-{ \tilde{\bf k}}_{\tau}^{2}$,  
${\tilde E}_{\nu}^{2}={\tilde{\bf q}}_{\nu}^{2}$  (the tilde refers to the laboratory frame), 
and ${ \tilde{\bf k}}_{\tau}=\kappa \bf{\hat x}$, where  $\bf{\hat x}$ is the unit 
vector along the line connecting the $\tau$ production and decay vertex,  
the factor $\kappa$ is obtained by solving this system of equations. For each  
$\tau$ lepton, we obtain two solutions, and we select the solution for which 
the sum of the transverse $\tau^\mp$ momenta is closest to zero. With  
this solution for ${\tilde k}_{\tau}^{\mu \mp}$ the $\tau^\mp$ rest frames and the 
momentum directions $\bf{\hat{q}}_{1,2}$ can be reconstructed, which are 
required for the  observables (\ref{bbzoadi}), (\ref{bbzqcp}), and~(\ref{bbzasq}). 
 
Fig. 1a shows the $\cos\varphi$ distributions for the production of a scalar 
$\phi =H$ and a pseudoscalar $\phi=A$ in the decay channel (\ref{bbzreac2}), assuming 
a mass $m_{H,A}=200$ GeV, both for no detector cuts and for applying 
the cuts 
$p_T\geq 40$~GeV and $\eta\geq 2.5$ (pseudorapidity) on the pions in 
the final state. 
In fact, the cut on $\eta$ does not change the shape of the normalized 
distributions shown in Fig.~1. 
The figure shows that the $p_T$ cuts have only a very minor influence, 
too.  
The slopes are given to very good approximation  by the numbers below 
in~(\ref{bbzoadi}). This implies that the shape of these distributions 
will be quite stable with respect to inclusion of higher order QCD corrections. 
The 
influence of the cuts on the shape of the distributions decreases for larger  
$\phi$ masses. Only for light Higgs masses  $m_{\phi}\gtrsim 120$ GeV 
does the chosen 
minimum $p_T$ cut of $40$ GeV have  a more significant effect. 
\begin{figure}[h] 
\begin{centering} 
\includegraphics[clip,scale=0.42]{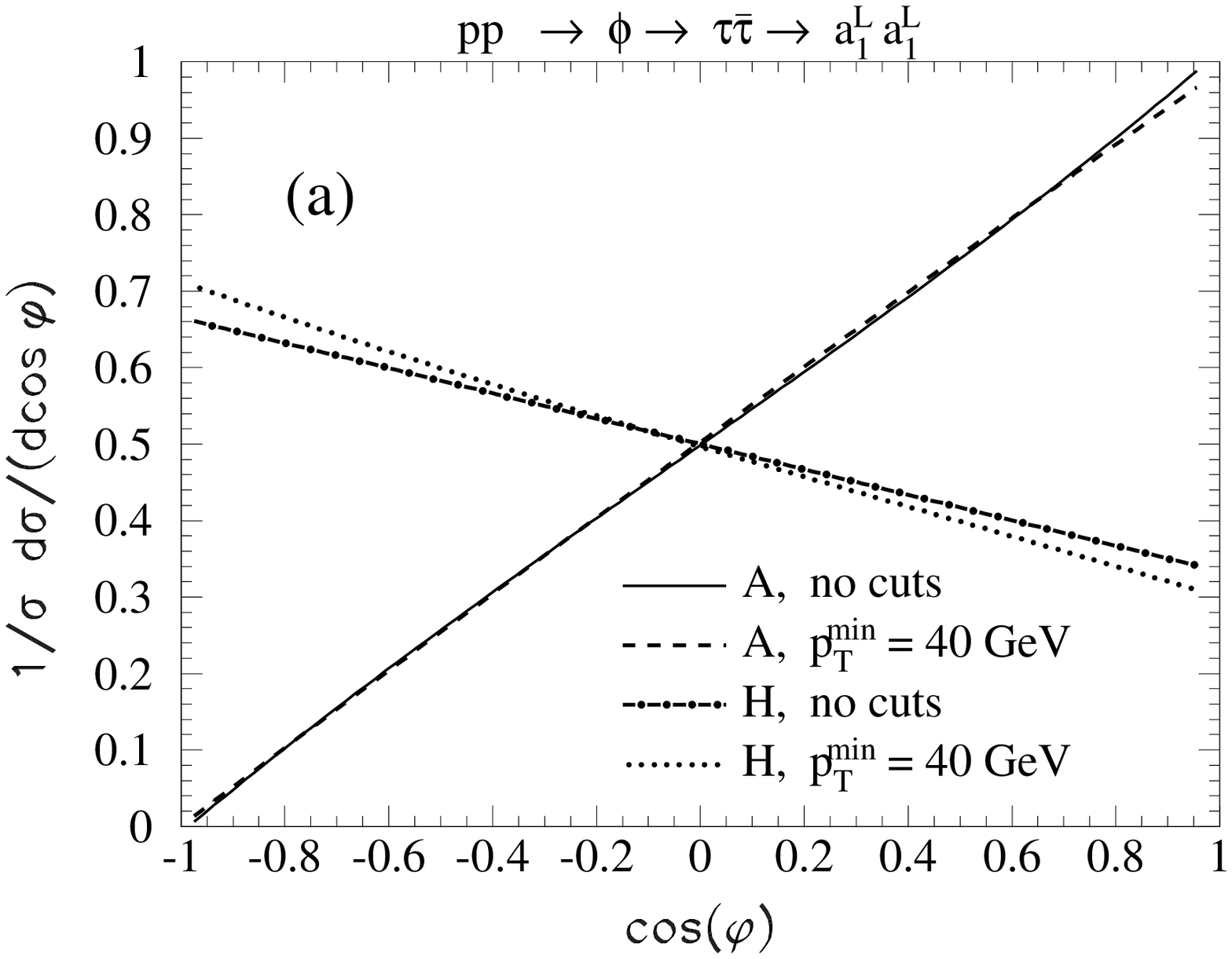}\hspace*{.4cm} 
\includegraphics[clip,scale=0.42]{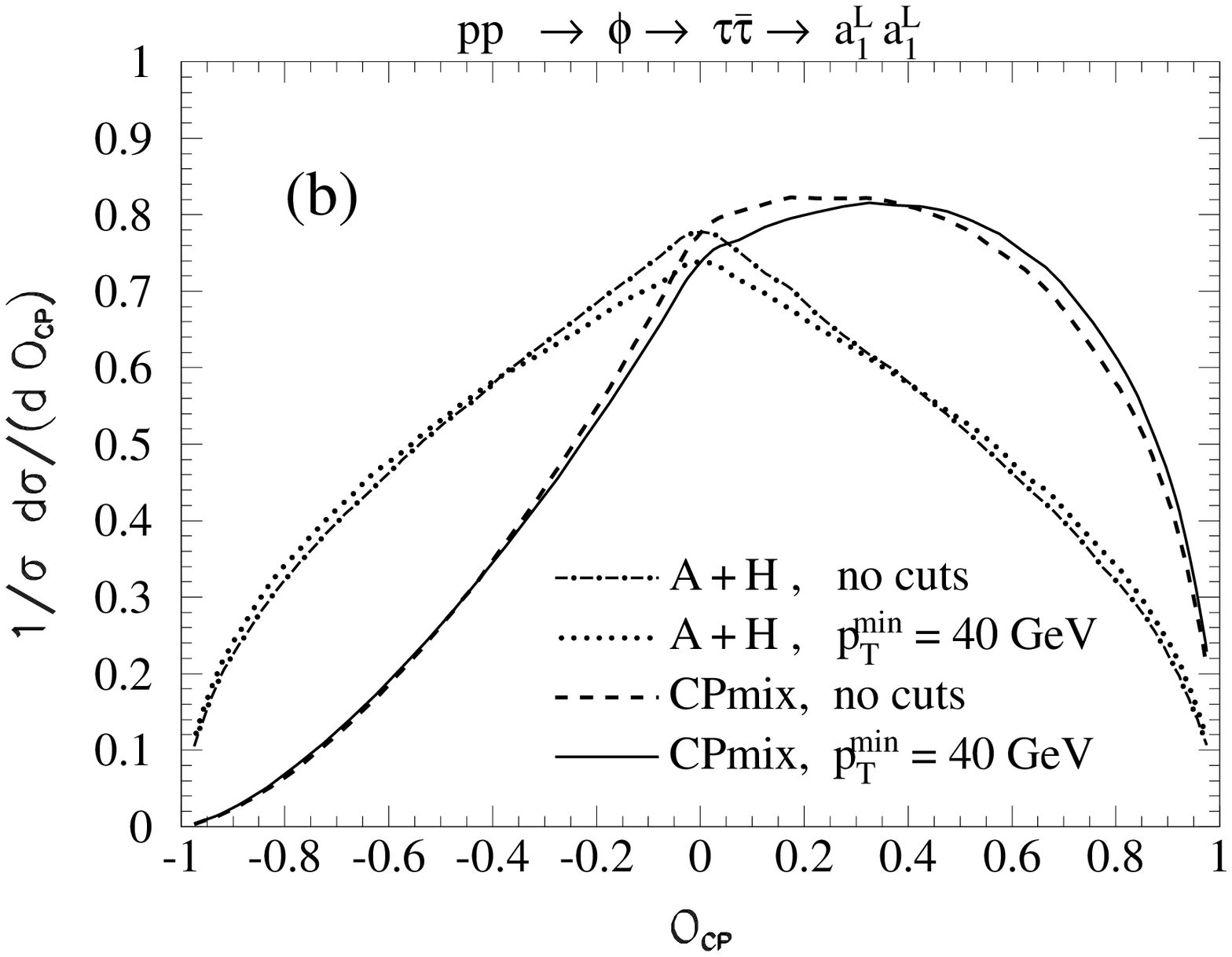}\vspace{-3pt} 
\par\end{centering} 
 
\caption{Distributions of $\cos\varphi$ (a) and ${\cal O}_{CP}$ (b) 
for  $m_\phi=200$~GeV and $a_{1}$ polarizations $LL$ or $TT$. 
\label{fig:TT_detectorcuts_mh200}} 
\end{figure} 
 
Let us now discuss the following two 
situations: i) Suppose both a scalar and a pseudoscalar 
Higgs boson with (nearly) degenerate masses, for instance  
$m_{H,A}\sim 200$ GeV, exist and are produced in the reaction  
(\ref{bbzreac2}), $i \to H, \, A$. Such degenerate resonances cannot be resolved, 
e.g.,  in the  $M_{\tau\bar \tau}$ spectrum. The resulting  
$\cos\varphi$ distribution will have a shape somewhere between the scalar and 
pseudoscalar extremes shown in Fig.~1a, depending on the relative reaction rates. 
ii) Suppose, on the other hand, that a Higgs boson $\phi$ with  
$m_{\phi}\sim 200$~GeV exists\footnote{Of course, Higgs bosons of this 
type might also be degenerate; for simplicity we do not consider this 
possibility here.} 
 which has both scalar and pseudoscalar couplings 
to fermions, in particular to $\tau$ leptons. 
The slope of the resulting $\cos\varphi$ distribution will also differ from the two extremes 
shown in Fig.~1a. In other words, the measured distribution does not tell 
whether degenerate scalar and pseudoscalar resonances 
or a state of undefined $CP$ parity were produced. This puzzle may be resolved 
using the observable ${\cal O}_{CP}$. As case i) corresponds to a $CP$-invariant Higgs sector, 
the resulting distribution of ${\cal O}_{CP}$ must be symmetric (if the phase space cuts 
are $CP$-symmetric) and $\langle {\cal O}_{CP}\rangle = 0$, while case ii) will 
produce an asymmetric distribution and a non-zero average. This is shown in 
Fig.~1b, where  case ii) is illustrated with  
 an ``ideal mixture'' (label $CPmix$), i.e., a $\phi$ boson 
with scalar and pseudoscalar couplings of equal magnitude -- we put
$a_\tau = - b_\tau$. 
Again, the applied cuts have only a minor influence on the 
distributions. 
As already mentioned above the 
distributions in Figs.~1a,~b do not change if both intermediate $a_1$ are transversely polarized, 
 while they are reflected with respect to the vertical line  
passing the abscissa value zero in the case of mixed polarizations. 
 
An important question is how robust is the discriminating power of these distributions 
with respect to experimental errors. In order to study this issue, using Monte Carlo 
methods, we have  
accounted for the  expected measurement uncertainties by ``smearing''  the relevant 
quantities with a Gaussian according to  
$\exp(-(x/\sigma)^2/2)$,  
where $x$ denotes the generated quantity (position in $x$ space, momentum component, 
energy) and $\sigma$ its expected standard deviation (s.d.).  
We use here 
$\sigma_{z}^{P}=15\,\mu m,\; \sigma_{T}^{S}=15\,\mu m,\; 
\sigma_{L}^{S}=500\,\mu m, \; \sigma_{\theta}^{a_{1}}=0.8\,{\rm 
  mrad},\; \delta E/E=2\%$,  
where  $\sigma_{z}^{P}$ denotes the s.d. of the position of the 
$\tau$ production vertex along the beam axis, while $\sigma_{L}^{S}$ 
and $\sigma_{T}^{S}$ are the s.d.  of the positions of the respective 
$\tau$ decay vertex along the $\tau$-jet axis (i.e., the direction of $a_1$) 
and in the plane transverse to this  axis. Furthermore, the uncertainty in determining the  
direction of $a_1$ is parameterized by an angle $\theta$ with s.d.  
$\sigma_{\theta}^{a_{1}}$,  
and $\delta E/E$ denotes the relative error of determining  
the energy of $a_{1}$. These values appear to be realistic for the 
LHC experiments \cite{genn,Tarrade:2007zz}. For this simulation  we 
use, in the case of $m_\phi=200$~GeV,  a constant $\tau$ flight length of $4.5$ mm.  
 
The effect of these uncertainties on the distributions of  
$\cos\varphi$ and ${\cal O}_{CP}$ is shown in Figs. 2a, b, using again Higgs 
boson masses $m_\phi = 200$ GeV.  
For the above set of uncertainties, scalar and pseudoscalar states are still clearly 
distinguishable (Fig. 2a), and likewise,  $CP$-conserving and 
$CP$-violating states  (Fig. 2b). 
\begin{figure}[h] 
\begin{centering} 
\includegraphics[clip,scale=0.42]{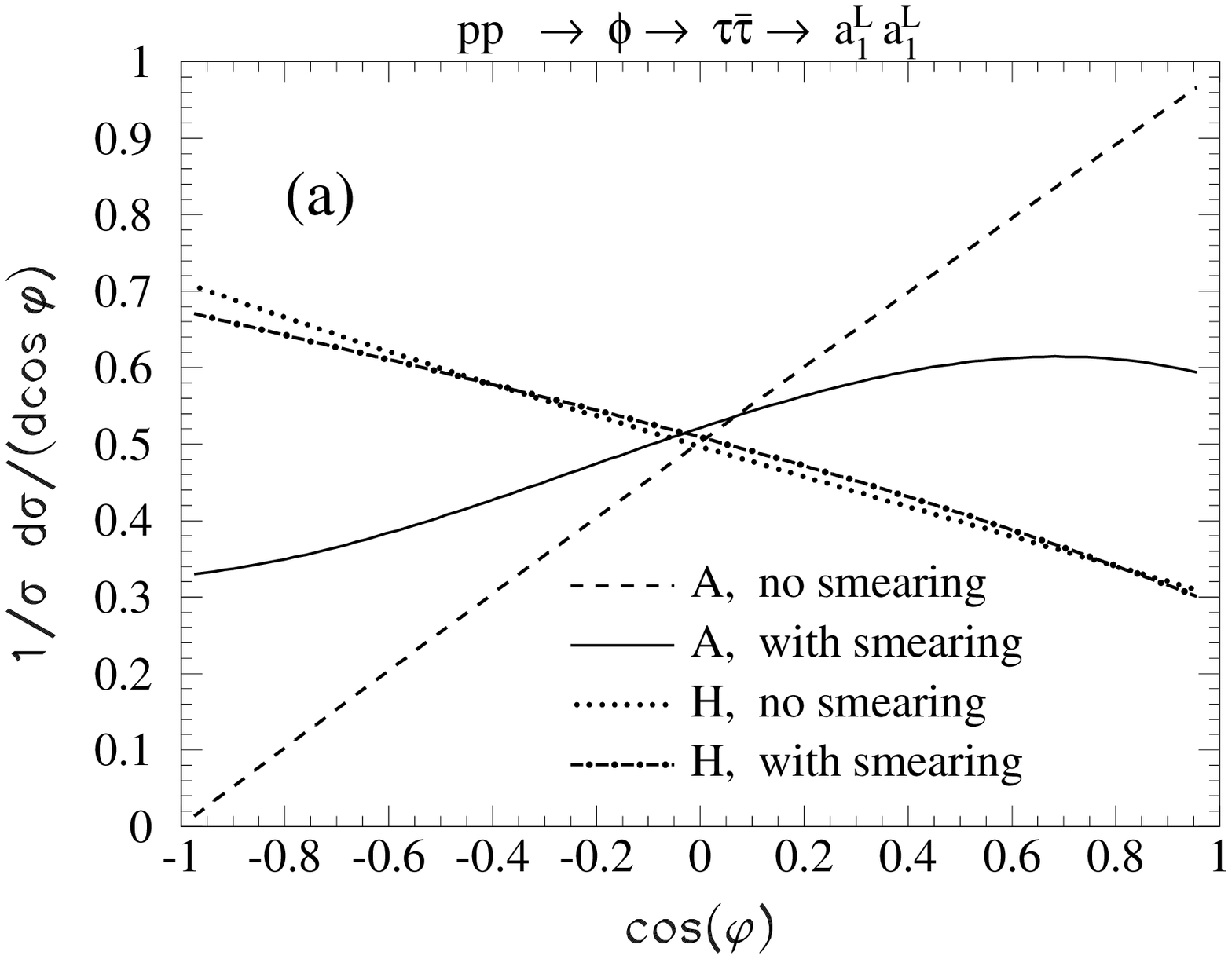}\hspace*{.4cm}%
\includegraphics[clip,scale=0.42]{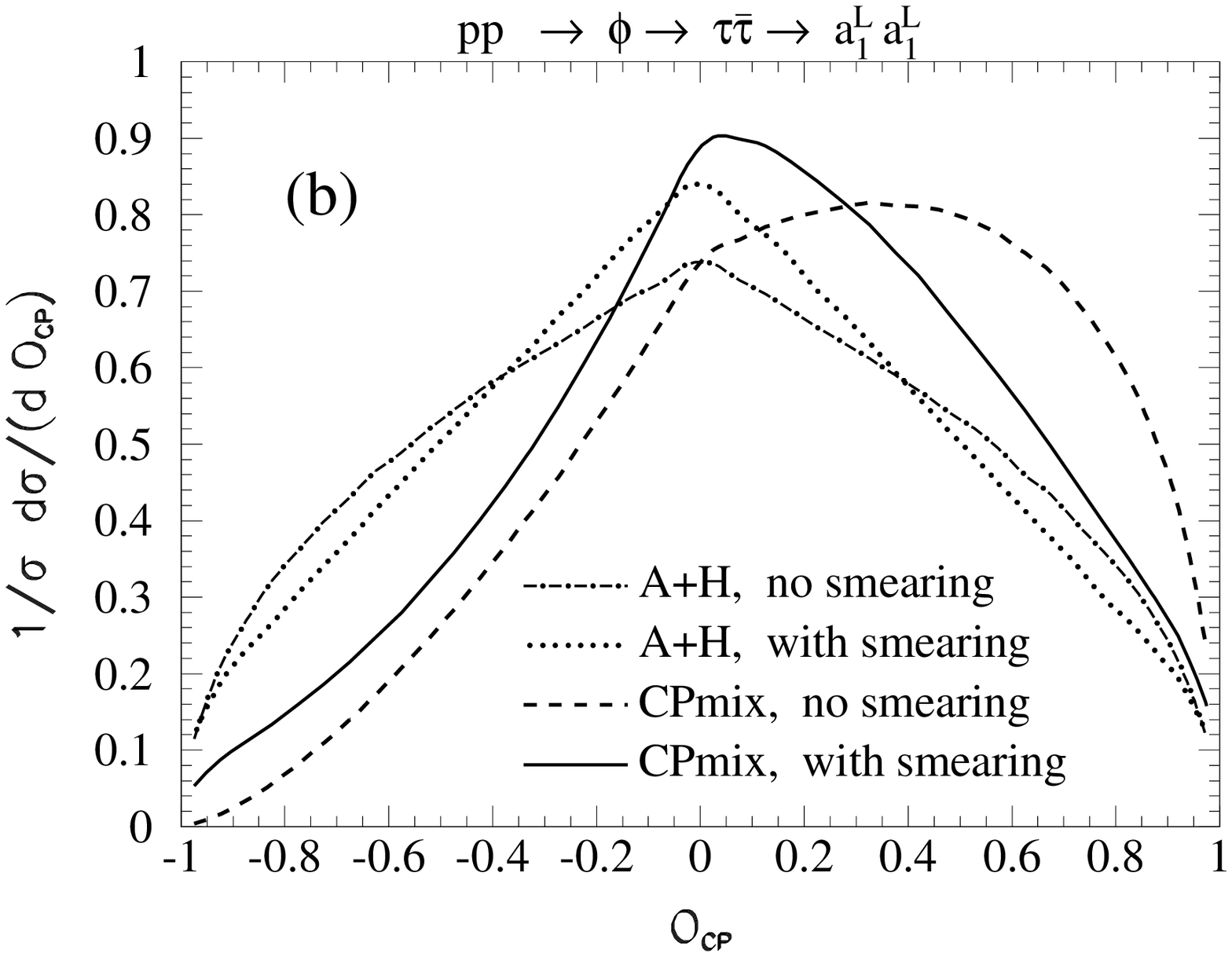}\vspace{-3pt} 
\par 
\end{centering} 
 
\caption{Distributions of $\cos\varphi$ (a) and ${\cal O}_{CP}$ (b) 
  taking into account measurement uncertainties and cuts. 
\label{fig:Smeared_mh200}} 
\end{figure} 
We have made a systematic study by 
varying i) the masses $m_\phi$ of the various types of Higgs bosons between 120 
GeV and 500 GeV and ii) the expected measurement errors.   
Varying $m_\phi$ we found that the discriminating power 
of these distributions does not decrease for heavy Higgs bosons. 
This can be understood as follows. When $m_\phi$ is increased the 
angle between the $\tau$ and $a_1$ directions in the laboratory frame 
decreases, which implies that the dependence on the smearing 
parameters  of the distributions is becoming stronger.  
On the other hand, the fact that  the average flight length of 
the $\tau$ leptons is becoming larger reduces the sensitivity 
to the smearing parameters,  
leaving the overall dependence of the distributions  on the above set of 
uncertainties rather stable. 
Concerning measurement errors we found that it is important to have 
under control 
the transverse uncertainty $\sigma_{T}^{S}$ in the reconstruction of 
the $\tau$ decay vertices and also the uncertainties $\sigma_{z}^{P}$ 
and $\sigma_{\theta}^{a_{1}}$ of the position of the  $\tau$ 
production vertex and of $\theta$. In order to 
make use of the discriminating power of the above observables, 
one should achieve  
$\sigma_{T}^{S} < 18\,\mu m$, 
 $\sigma_{z}^{P}< 30\,\mu m$, and   
$\sigma_{\theta}^{a_{1}} < 1 \,{\rm mrad}$ in future experiments. 
Least critical are the resolution of the longitudinal $\tau$-jet 
axes ($\sigma_{L}^{S}$) and the energy uncertainty of the $a_{1}$ meson. 
Details of our results will be given elsewhere~\cite{bebe08}. 
 
Finally, we estimate how many events (\ref{bbzreac2}) 
are needed in order to discriminate between i) a scalar and pseudoscalar 
Higgs boson and/or ii) between $CP$-conserving and 
$CP$-violating states, assuming $m_\phi = 200$~GeV.  
As to i),  we define an asymmetry $A_\varphi =  
[N(\cos\varphi>0) - N(\cos\varphi<0)]/[N_> +N_<]$.  
From Fig. 2a we obtain from the smeared distributions 
 $A_{\varphi}^H = -0.19$ and  
$A_{\varphi}^A = 0.17$. Thus, for distinguishing $H$ from $A$ with 
3 s.d. significance requires 69 events  (\ref{bbzreac2}). 
Concerning ii), the result of Fig. 2b implies that for an ideal $CP$ 
mixture the CP asymmetry (\ref{bbzresa}) takes the value 
$A({\cal O}_{CP}) = 0.23$ while it is  zero for pure $H$, $A$, and  
degenerate $H$ and $A$ intermediate states. Thus, 170 events 
(\ref{bbzreac2}) will be needed  to establish this CPV effect at the 3 s.d. level. 
This may be feasible, depending on the masses and couplings of 
$\phi$, after several years of high luminosity runs at  
the LHC \cite{:1999fr,Ball:2007zza}.  
The $\tau$ data sample will be increased by an order of magnitude if the 
above observables can be applied also to one-prong hadronic $\tau$ 
decays. This is currently being investigated, 
employing appropriately constructed pseudo rest-frames \cite{bebe08}.\\[1pt]

{\it \bf Conclusions:} \quad The $\tau$ decay channel is clearly most suited to 
 explore the $CP$ nature of a light or heavy neutral Higgs boson $\phi$. This 
 is an important physics issue if Higgs bosons are discovered.  
We have discussed a set of observables that serve this purpose, and we have 
shown, for Higgs boson production at the LHC  
and its decay via $\tau$ leptons into $a_1$ mesons, that the above 
correlations and asymmetries provide powerful tools for 
discriminating  between $CP$-even and 
-odd Higgs bosons and for searches for $CP$ violation in the Higgs 
sector. The measurement of these observables is challenging, but our 
analysis indicates that it should be feasible in the long run, 
provided enough $\phi \to \tau \tau$ events will be recorded at the LHC.  \\ \\ 
%
{\it \bf Acknowledgements:} \quad We thank  P. Sauerland and A. Stahl for helpful discussions. 
This work was supported by Deutsche Forschungsgemeinschaft  
 SFB/TR9.
\newpage



\begin{thebibliography}{99} 
 
\bibitem{Djouadi:2005gi} 
  A.~Djouadi, 
   arXiv:hep-ph/0503172; arXiv:hep-ph/0503173. 
 
 
\bibitem{Accomando:2006ga} 
  E.~Accomando {\it et al.}, 
  arXiv:hep-ph/0608079. 
 
\bibitem{Cohen:1993nk} 
  A.~G.~Cohen, D.~B.~Kaplan and A.~E.~Nelson, 
  Ann.\ Rev.\ Nucl.\ Part.\ Sci.\  {\bf 43}, 27 (1993). 
 
 
\bibitem{Bernreuther:1993df} 
  W.~Bernreuther and A.~Brandenburg, 
  Phys.\ Lett.\  B {\bf 314}, 104 (1993); Phys.\ Rev.\  D {\bf 49}, 4481 (1994). 
 
\bibitem{Bernreuther:1997af} 
  W.~Bernreuther, A.~Brandenburg and M.~Flesch, 
  Phys.\ Rev.\  D {\bf 56}, 90 (1997). 
 
\bibitem{Bernreuther:1998qv} 
  W.~Bernreuther, A.~Brandenburg and M.~Flesch, 
  arXiv:hep-ph/9812387. 
 
 
\bibitem{Dell'Aquila:1988fe} 
  J.~R.~Dell'Aquila and C.~A.~Nelson, 
  Nucl.\ Phys.\  B {\bf 320}, 86 (1989). 
 
 
\bibitem{Chang:1993jy} 
  D.~Chang, W.~Y.~Keung and I.~Phillips, 
  Phys.\ Rev.\  D {\bf 48}, 3225 (1993). 
 
\bibitem{Kramer:1993jn} 
  M.~Kr\"amer, J.~H.~K\"uhn, M.~L.~Stong and P.~M.~Zerwas, 
  Z.\ Phys.\  C {\bf 64}, 21 (1994). 
 
 
\bibitem{Grzadkowski:1995rx} 
  B.~Grzadkowski and J.~F.~Gunion, 
  Phys.\ Lett.\  B {\bf 350}, 218 (1995). 
 
\bibitem{Bernreuther:1992dz} 
  W.~Bernreuther, T.~Schr\"oder and T.~N.~Pham, 
  Phys.\ Lett.\  B {\bf 279}, 389 (1992). 
 
\bibitem{Dell'Aquila:1985ve} 
  J.~R.~Dell'Aquila and C.~A.~Nelson, 
  Phys.\ Rev.\  D {\bf 33}, 80 (1986); 
  T.~Plehn, D.~L.~Rainwater and D.~Zeppenfeld, 
  Phys.\ Rev.\ Lett.\  {\bf 88}, 051801 (2002); 
  C.~P.~Buszello {\it et al.}, 
  Eur.\ Phys.\ J.\  C {\bf 32}, 209 (2004). 
 
\bibitem{Stahl:2000aq} 
  A.~Stahl, 
  Springer Tracts Mod.\ Phys.\  {\bf 160}, 1 (2000). 
 
\bibitem{Rouge:1990kv} 
  A.~Roug{\'e}, 
  Z.\ Phys.\  C {\bf 48}, 75 (1990). 
 
\bibitem{Davier:1992nw} 
  M.~Davier, L.~Duflot, F.~Le Diberder and A.~Rou{g\'e}, 
  Phys.\ Lett.\  B {\bf 306}, 411 (1993). 
 
\bibitem{Kuhn:1995nn} 
  J.~H.~K\"uhn, 
  Phys.\ Rev.\  D {\bf 52}, 3128 (1995). 
 
\bibitem{:1999fr} 
  ``ATLAS detector and physics performance.  
  Technical design report.  Vol. 2,'', report CERN-LHCC-99-15. 
 
 
\bibitem{Ball:2007zza} 
  G.~L.~Bayatian {\it et al.}  [CMS Collaboration], 
  J.\ Phys.\ G {\bf 34}, 995 (2007). 
 
 
\bibitem{genn} 
  S.~Gennai {\it et al.}  [CMS Collaboration], 
  Eur.\ Phys.\ J. {\bf C46, \, S01}, 1 (2006). 
   
\bibitem{Tarrade:2007zz} 
  F.~Tarrade  [ATLAS Collaboration], 
  Nucl.\ Phys.\ Proc.\ Suppl.\  {\bf 169}, 357 (2007). 
 
 
 
 
 
\bibitem{bebe08} 
S. Berge and W. Bernreuther, to be published. 
 
\end{thebibliography}
\end{document}